\documentclass[pre,aps,twocolumn,showpacs,amsmath,amssymb,amsfonts]{revtex4}
\usepackage{epsfig}
\begin{document}
\title{Crossover from the pair contact process with diffusion
to directed percolation}
\author{Su-Chan Park}
\affiliation{School of Physics, Korea Institute for Advanced Study, Seoul
130-722, Korea}
\author{Hyunggyu Park}
\affiliation{School of Physics, Korea Institute for Advanced Study, Seoul
130-722, Korea}
\date{\today}
\begin{abstract}
Crossover behaviors from the pair contact process with diffusion (PCPD) and the
driven PCPD (DPCPD) to the directed percolation (DP) are studied in one
dimension by introducing a single particle annihilation and/or 
branching dynamics. The
crossover exponents $\phi$ are estimated numerically as $1/\phi \simeq
0.58\pm0.03$ for the PCPD and $1/\phi \simeq 0.49 \pm 0.02$ for the DPCPD.
Nontriviality  of the PCPD crossover exponent strongly 
supports  the non-DP nature
of the PCPD critical scaling, which is further evidenced by the anomalous
critical amplitude scaling near the PCPD point. In addition, we find that the
DPCPD crossover is consistent with the mean field prediction of the tricritical
DP class as expected.
\end{abstract}
 \maketitle 
The absorbing phase transition
(APT) has emerged during the last few decades as a prototype of
nonequilibrium critical phenomena. The APT is a transition from an active phase
into an inactive (absorbing) phase which is composed of absorbing (trapped)
configurational states where the system cannot leave by the prescribed dynamic
rules.  As in the equilibrium critical phenomena, a few scaling exponents
characterize and classify critical behaviors into universality classes
\cite{H00,O04}. Robustness of the directed percolation (DP) universality class
with respect to the microscopic details led to the ``DP conjecture''
\cite{J81,G82} that a model should belong to the DP class if it has a unique
absorbing state without additional symmetry, conservation laws, quenched
disorder, and long-range interactions. Although significant progress has
been achieved to date, the full understanding on the main features of the APT
universality classes is still far from complete.

To achieve such a goal, it is crucial to identify the universality class of the
pair contact process with diffusion (PCPD) \cite{G82,HT97,CHS01}, which has
been the most controversial topic these days, to our knowledge. 
The PCPD is a variant of the pair
contact process \cite{J93} by allowing single-particle diffusion in addition to
binary fission and annihilation dynamics 
($2A \rightarrow 3A$ and $2A \rightarrow
\emptyset$). Even with combined efforts of extensive and highly equipped
numerical and analytical studies, its one-dimensional version has defied a
consensus as yet \cite{HH04}. Rather, various possible scenarios have been
suggested with the DP class in the center of this controversy. In one scenario,
the PCPD should belong to a universality class distinct from the DP with
a unique set of critical exponents \cite{KC03,PP05a,PP05b} or continuously
varying exponents due to the marginal perturbation \cite{NP04}. The other
states that the PCPD will eventually flow to the DP fixed point after a huge
crossover time \cite{H03,BC03,H05}. More scenarios can be found in
Ref.~\cite{HH04}. In high dimensions, it is unquestionably clear that the PCPD
and the DP exhibit different critical scaling.

The strong corrections to scaling are the main obstacle for the numerical study
to determine the universality class of the PCPD. To make matters worse, the
field theory with a single component order parameter is shown to be not viable
\cite{JvWOT04}. It implies that the proper field theory, if it exists, needs (at
least) two independent order parameters, which is also independently proposed
by us from the numerical study of the driven PCPD (DPCPD)\cite{PP05a}.

\begin{figure}[b]
\includegraphics[width=0.42\textwidth]{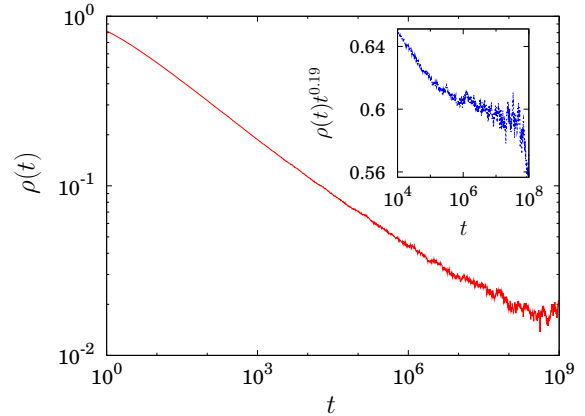}
\caption{\label{Fig:PCPD_pc} (Color online) Log-log plot of the particle
density decay for the PCPD at $p=0.133~516$. After $t\simeq 10^8$, the density
saturates, which signals that the system is in the active phase. Inset:
Semilogarithmic plot of $\rho(t) t^{\delta}$ versus $t$ at
$p=0.133~522$ with $\delta = 0.19$.
After $t\simeq 10^7$, the curve veers down, which signals that
the system is in the absorbing phase. This result is insensitive to the change
of the $\delta$ value by up to $15\%$.}.
\end{figure}
In this Rapid Communication, we propose an approach to settle down or at least moderate
the PCPD controversy by scrutinizing closely the crossover behavior from the
PCPD to the DP class. We allow unary branching and/or annihilation dynamics ($A
\rightarrow 2A$ and $A \rightarrow \emptyset$) in the PCPD model, which should
lead to the well-known DP critical dynamics as soon as it is introduced. The 
{\em nontrivial} nature of the DP critical line approaching the PCPD point as well
as {\em anomalous} crossover scaling behavior would serve as an indicative of
different scaling at the PCPD point, distinct from the DP.

Although the accurate estimation of the critical exponent values for the PCPD
is severely plagued by corrections to scaling, the critical point itself can be
located with relatively high precision. In Fig. \ref{Fig:PCPD_pc}, the density
decay near criticality is shown for the PCPD model of which detailed dynamics
is given in Eq.~\eqref{Eq:PCPD} with $w=0$. The system size at a tuning
parameter $p=0.133~516$ (0.133 522) is $2^{18}$ ($2^{20}$) and the number of
independent samples is 32 (104). All samples carry at least one neighboring
pair until the end of simulations, $t=10^9~(10^8)$, which guarantees that
finite size effects have not emerged as yet. From Fig.~\ref{Fig:PCPD_pc}, the
critical point is estimated as $p_c = 0.133~519 (3)$ with the number in the
parentheses as the uncertainty of the last digit. It should be emphasized that
our estimation of the critical point for the PCPD does not resort to any
prescribed critical exponent value.

In the generalized crossover model including unary branching and/or annihilation
processes, it is relatively easier to locate the critical points
accurately by utilizing the known DP critical exponent values. With the
annihilation $A\rightarrow \emptyset$ and/or the branching
$A\rightarrow 2A$, the density decay in the absorbing phase should be
exponential in time at least in one dimension, which nullifies the long-term
memory effect of the isolated particles \cite{NP04} and the simple unplagued DP
scaling should be anticipated. The accurate information on the critical line as
a function of the unary process rate $w$ allows us to estimate the crossover
exponent $\phi$, which describes the crossover scaling, if it exists, from the
PCPD to the DP.

The detailed evolution rules for the generalized crossover model are summarized
using stoichiometric notations as
\begin{subequations}
\label{Eq:PCPD}
\begin{eqnarray}
&&AA \rightarrow \emptyset \emptyset \text{ with rate $\lambda$},\label{Eq:PCPDa}\\
&&\left .\begin{matrix}AA\emptyset\\\emptyset AA \end{matrix}
\right \} \rightarrow AAA, \text{ with rate $\sigma/2$} \label{Eq:PCPDb},\\
 &&A\emptyset \rightarrow \left \{
\begin{matrix} \emptyset A & \text{with rate $D (1-w)/2$}\\
\emptyset\emptyset &\text{with rate $w(1-q)/2$}\\
A A &\text{with rate $wq/2$} \end{matrix} \right .
\label{Eq:PCPDc},\\
 &&\emptyset A\rightarrow \left \{
\begin{matrix} A \emptyset  & \text{with rate $D (1-w)/2$}\\
\emptyset\emptyset &\text{with rate $w(1-q)/2$}\\
A A &\text{with rate $w q/2$} \end{matrix} \right .
\label{Eq:PCPDd},
\end{eqnarray}
\end{subequations}
where $A$ ($\emptyset$) stands for a hard core particle (vacancy) and $0
\le q \le 1$. The periodic boundary conditions are employed on a one-dimensional
lattice with size $L$. The case of $w=0$, $D=1$, $\lambda =p$, and $\sigma=1-p$
with a tuning parameter $p$ corresponds to the PCPD model studied before
\cite{PP05a}, of which the critical point is accurately located through Fig.
\ref{Fig:PCPD_pc}. In the case of nonzero $w$, the critical point is located by
observing the flatness of $\rho(t) t^{\delta_\text{DP}}$
as a function of $t$ with the DP
critical exponent $\delta_\text{DP} \simeq 0.1595$ \cite{J96}. In the following,
we always set $D=1$ and $\lambda = 1-\sigma =p$.

The simulation algorithm to mimic the dynamics in Eq. \eqref{Eq:PCPD} is as
follows: First, choose a particle at random. Then, choose one of its nearest
neighboring sites randomly as a target site. If the target site is vacant, the
chosen particle is annihilated with probability $w(1-q)$, branches a particle
at the target site with probability $w q$, or hops to that site with
probability $1-w$. The random selection of one of nearest neighbors amounts to
the factor $\frac{1}{2}$ of the transition rates in Eqs. \eqref{Eq:PCPDc} and
\eqref{Eq:PCPDd}. In the case where the target site is already occupied, both
particles are annihilated with probability $p$ or one extra particle is
generated at a randomly chosen nearest neighbor site of the pair with
probability $1-p$ when that site is vacant. If the chosen site is already
occupied, this branching attempt is rejected. After this update, the time
increases by $1/N_t$, where $N_t$ is the total number of particles at time $t$
just before the update.

\begin{table}[t]
\caption{\label{Table1} Critical point values $p_{c0}(w)$ and $p_{c1} (w)$ for
various $w$'s for $q = 0$ and $q=1$, respectively. The numbers in the
parentheses indicate the uncertainty of the last digits.}
\begin{ruledtabular}
\begin{tabular}{rll}
$w$&$p_{c0}(w)$&$p_{c1}(w)$\\
\hline
0  &0.133~519(3) &0.133~519(3)\\
$10^{-5}$&0.133~172(3)&0.134~085(5)\\
$5\times 10^{-5}$&0.132~581(2) &\\
$10^{-4}$&0.132~080(3)&0.135~81(1)\\
$2\times 10^{-4}$&0.131~325(5) &0.136~98(1)\\
$3\times 10^{-4}$&0.130~715(5)&0.137~92(1)\\
$4\times 10^{-4}$&0.130~185(5) &0.138~73(1)\\
$5\times 10^{-4}$&0.129~708(3)&0.139~46(1)
\end{tabular}
\end{ruledtabular}
\end{table}
\begin{figure}[b]
\includegraphics[width=0.42\textwidth]{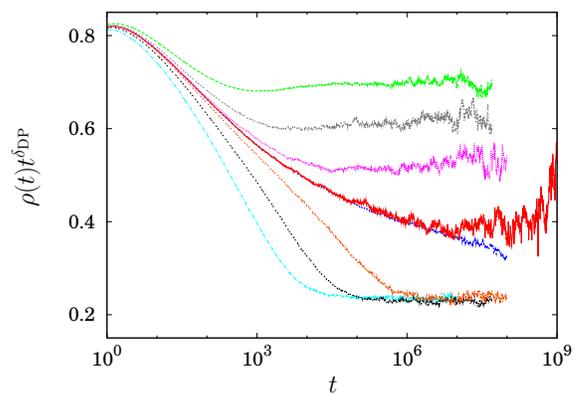}
\caption{\label{Fig:amplitude}(Color online) Semilogarithmic plots of $\rho(t)
t^{\delta_\text{DP}}$ as a function of $t$
for the generalized crossover model at
criticality. The upper (lower) three curves correspond to $w=5\times
10^{-4},~10^{-4}$, and $10^{-5}$ from top (left) to bottom (right) at $q=0$
($1$). The middle two curves are redrawn using the same data as in Fig.
\ref{Fig:PCPD_pc} for the PCPD ($w=0$) near the critical point. }
\end{figure}
By observing how $\rho(t) t^{\delta_\text{DP}}$ behaves in the asymptotic
regime, the critical point values $p_c$ for nonzero $w$'s are estimated. If the
system is in the active (absorbing) phase, those curves should veer up (down).
Only at criticality, a flat straight line can be observed. Initially, all sites
are occupied. The simulations were performed up to $t = 10^7 - 10^8$
and no finite size effects have been observed with system size $L=2^{18}
- 2^{20}$. The
numerical estimates for the critical point values $p_{c0}(w)$ at $q = 0$ and
$p_{c1} (w)$ at $q=1$ are tabulated in Table \ref{Table1}. Figure
\ref{Fig:amplitude} shows the critical density decay for various values of $w$
at $q=0$ and 1. The flatness of the critical curves can be seen quite early
even for very small $w$, while the PCPD lines ($w=0$) do not show any flatness
within our observation time. It may be understood either by a tremendously long
dynamic correction to the DP scaling \cite{H05} or by a non-DP scaling for the
PCPD. In either case, it is evident that there exists a diverging crossover
time scale as $w\rightarrow 0$.
\begin{figure}[t]
\includegraphics[width=0.42\textwidth]{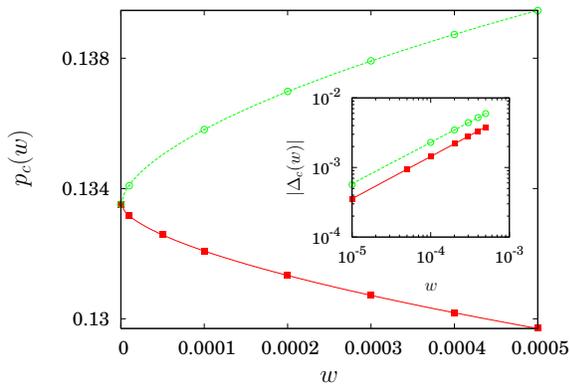}
\caption{\label{Fig:phi} (Color online) Critical points $p_c(w)$ for various
$w$'s and their fitting functions. The lower (upper) curve corresponds to $q=0$
($1$). The error bar is smaller than the symbol size. The curves represent the
fitting results of the critical points with the function $\Delta_c(w)= a
w^{1/\phi}$. The estimated crossover exponent is $1/\phi = 0.58 \pm 0.03$.
Inset: Log-log plot of $|\Delta_c(w)|$ vs $w$. The slope of the straight line
is $1/\phi$. }
\end{figure}

The crossover scaling function near the PCPD critical point  should take the
form
\begin{equation}
\rho(w,\Delta;t) = t^{-\delta} F(\Delta w^{-1/\phi}, t w^{\mu_\|} ),
\label{Eq:CrossOver}
\end{equation}
where $\mu_\| = \nu_\|/\phi$ with the crossover exponent $\phi$, $\nu_\|$, and
$\delta$ are the critical exponents of the PCPD, and $\Delta = p - p_c(0)$ with
$p_c(0)$ being the PCPD critical point value. The crossover time diverges as
$w\rightarrow 0$ such that  $\tau_\text{cross}\sim w^{-\mu_\|}$.

The crossover exponent $\phi$ can be calculated without the knowledge of the
values of $\nu_\|$ and $\delta$ by studying how the DP critical line $p_c(w)$
approaches the PCPD critical point. This line should be one of the
renormalization group flow lines such as $\Delta_c(w) \simeq a w^{1/\phi}$
\cite{DL9}, where $\Delta_c(w) \equiv p_c(w) - p_c(0)$
and $a$ is a (nonuniversal) constant.
In Fig. \ref{Fig:phi}, the crossover exponent is estimated using the
least-square fitting, which turns out to have a nontrivial and universal value
of $1/\phi = 0.58\pm 0.03$. This nontrivial nature of the critical line
($\phi\neq 1$) signals strongly a possible non-DP scaling at the PCPD point.

\begin{figure}[t]
\includegraphics[width=0.42\textwidth]{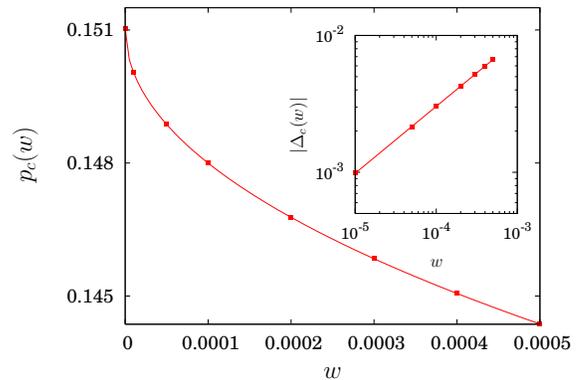}
\caption{\label{Fig:DPCPD} (Color online) Phase boundary of the DPCPD crossover
model with \eqref{Eq:PCPDcprime}. Just as in Fig. \ref{Fig:phi}, the
crossover exponent is estimated by the least-square fitting as $1/\phi = 0.49
\pm 0.02$ which is consistent with the mean field crossover exponent
\cite{Tri}. Inset: Log-log plot of $|\Delta_c(w)|$ vs $w$.  }
\end{figure}
\begin{figure}[b]
\includegraphics[width=0.42\textwidth]{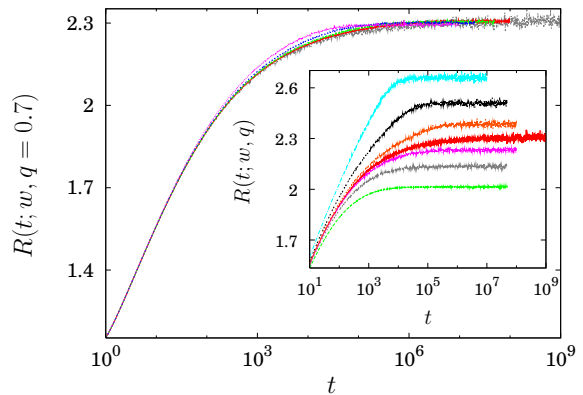}
\caption{\label{Fig:ratio} (Color online) Density-pair density ratio $R(t)$
for various $w$'s with $q=0.7$ in the semilogarithmic scale. The critical points are located at $p_c =
0.134~120(5)$, 0.133~761(2), 0.133~616(2), and 0.133~563(2) for
$w=10^{-4},~2\times10^{-5},~4\times 10^{-6}$, and $10^{-6}$, respectively. The
longest curve corresponds to the PCPD at $p=0.133~516$. The saturating values
are almost identical for all $w$'s including the PCPD. Inset: Plots of $R(t)$
for $w$'s in Fig. \ref{Fig:amplitude}. The upper (lower) three curves
corresponds to $q=1$ (0). For both cases, the ratio of two critical amplitudes
approaches to the PCPD value as $w$ becomes smaller. }
\end{figure}
The similar analysis can be applied to the  DPCPD with diffusion bias by
modifying the dynamics in Eqs. \eqref{Eq:PCPD} in such a way that the dynamics
\eqref{Eq:PCPDd}  is absent and the dynamics \eqref{Eq:PCPDc} is replaced by
\begin{equation}
A\emptyset \rightarrow \left \{
\begin{matrix} \emptyset A & \text{with rate $ (1-w)$}\\
\emptyset\emptyset &\text{with rate $w$}
\end{matrix} \right . .
\tag{$1c'$}
\label{Eq:PCPDcprime}
\end{equation}
With $w=0$, this model is the DPCPD studied in Ref. \cite{PP05a}. Since the
diffusion bias should not change the DP scaling \cite{PP05a}, this model with
nonzero $w$ is still expected to belong to the DP class. Figure \ref{Fig:DPCPD}
shows how the DPCPD crosses over to the DP as $w$ increases and finds the
crossover exponent $1/\phi \simeq 0.49\pm 0.02$, which is consistent with the
mean field prediction of the tricritical DP class \cite{Tri} and, in turn,
confirms again that the upper critical dimension of the DPCPD is 1
\cite{comment}.

To get a more concrete evidence for the non-DP nature of the PCPD scaling, we go
back to the scaling function in Eq. \eqref{Eq:CrossOver} which can be rewritten
at criticality as
\begin{equation}
t^{\delta_\text{DP}} \rho(w,\Delta_c;t) = t^{\delta_\text{DP} -\delta}
F(a,w^{\mu_\|} t) = w^\chi {\cal G} (w^{\mu_\|} t ),
\end{equation}
where $\chi = \mu_\|(\delta-\delta_\text{DP})$ and ${\cal G}(x) =
x^{\delta_\text{DP} - \delta} F(a,x)$. Since ${\cal G}(x)$ should be a finite
constant as  $x \rightarrow \infty$ to guarantee the DP scaling for nonzero
$w$, the critical decay amplitude is proportional to $w^\chi$ for sufficiently
small $w$ where the crossover scaling is valid.

If the PCPD belongs to the DP class ($\delta=\delta_\text{DP}$), the amplitude
converges to a nonzero value as $w\rightarrow 0$. Otherwise, it decreases
algebraically with $\chi\simeq 0.03$, using the present best estimates for the
PCPD exponent values as $\delta\simeq 0.19$ and $\nu_\|\simeq 1.85$
\cite{KC03,PP05a,H05}. In Fig.~\ref{Fig:amplitude}, there are apparently two
very different convergent behaviors for $q=0$ and 1. Naive estimates  lead to
$\chi\simeq 0.08$ for $q=0$ and $\chi\simeq 0.00$ for $q=1$, both of which do
not simultaneously fit into either the DP or the non-DP scenario.

\begin{figure}[t]
\includegraphics[width=0.42\textwidth]{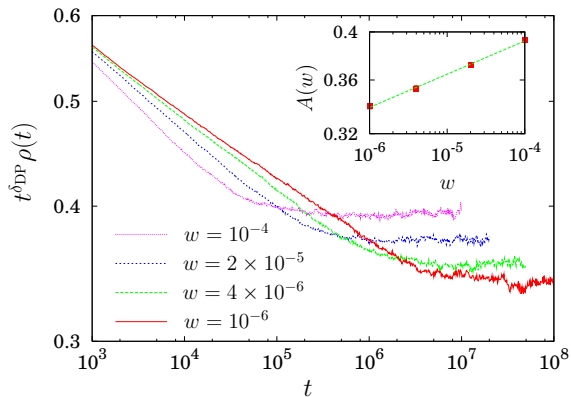}
\caption{\label{Fig:chi} (Color online) Log-log plots of $\rho(t)
t^{\delta_\text{DP}}$ vs $t$ for various $w$'s as in Fig. \ref{Fig:ratio}
with $q=0.7$ at criticality.
Inset: Log-log plot of the critical amplitudes $A(w)$ vs $w$. The slope of the
straight line is 0.03. }
\end{figure}
We find that this perplexing result originates from the narrowness of the
crossover scaling region near the PCPD point. To get a hint for the valid
scaling regime, we observe the pair density $\rho_p$ as well as the particle
density $\rho$ in simulations. Both quantities should scale in the same way and
their critical amplitudes also scale similarly as $\sim w^\chi$. Hence, the
ratio $R(t) \equiv \rho(t)/\rho_p(t)$ should be independent of $w$ for the
valid crossover scaling regime in the asymptotic limit. The inset of Fig.
\ref{Fig:ratio} shows that both cases at $q=0$ and 1 have not entered the
scaling regime as yet up to $w=10^{-5}$, which implies that the naive estimates
for $\chi$ should be significantly influenced by the corrections to scaling.
Fortunately, we find a reasonably good scaling regime for not so small $w$ at
$q=0.7$. Figure \ref{Fig:ratio} shows that the scaling regime is reached
already for $w=10^{-4}$. The inset of Fig. \ref{Fig:chi} shows how the critical
amplitude of the density, say $A(w)$, behaves in the crossover scaling regime,
from which we estimate $\chi = 0.03(1)$ that is consistent with the PCPD
estimate. This result strongly supports again the non-DP nature of the PCPD
scaling. The crossover exponent estimated for the case of $q=0.7$ is
$1/\phi\simeq 0.57$, which is compatible with the previous estimation within
error.

To conclude, we studied the crossover behavior from the pair contact process
with diffusion and the driven pair contact process with diffusion to the
directed percolation. We found that the crossover for the PCPD to the DP is
nontrivial and the critical amplitude scaling is anomalous, which strongly
supports the non-DP nature of the PCPD scaling. The crossover exponent for the
DPCPD takes the mean field value of the tricritical DP (TDP) \cite{Tri}. This
implies that the two-dimensional PCPD  might have connections to the TDP that
is under our current investigation.


\begin{thebibliography}{99}
\bibitem{H00} H. Hinrichsen, Adv. Phys. {\bf 49}, 815 (2000).
\bibitem{O04} G. \'Odor, Rev. Mod. Phys. {\bf 76}, 663 (2004).
\bibitem{J81} H.K. Janssen, Z. Phys. B: Condens. Matter {\bf 42}, 151 (1981).
\bibitem{G82} P. Grassberger, Z. Phys. B: Condens. Matter {\bf 47}, 365 (1982).
\bibitem{HT97} M.J. Howard and U.C. T\"auber, J. Phys. A {\bf 30}, 7721 (1997).
\bibitem{CHS01} E. Carlon, M. Henkel, and U. Schollw\"ock, Phys. Rev. E
{\bf 63}, 036101 (2001).
\bibitem{J93} I. Jensen, Phys. Rev. Lett. {\bf 70}, 1465 (1993).
\bibitem{HH04} For a review, see M. Henkel and H. Hinrichsen, J. Phys. A {\bf 37}, R117 (2004).
\bibitem{KC03} J. Kockelkoren and H. Chat\'e, Phys. Rev. Lett. {\bf 90},
125701 (2003).
\bibitem{PP05a} S.-C. Park and H. Park, Phys. Rev. Lett. {\bf 94},
065701 (2005).
\bibitem{PP05b} S.-C. Park and H. Park, Phys. Rev. E {\bf 71}, 016137 (2005).
\bibitem{NP04} J.D. Noh and H. Park, Phys. Rev. E {\bf 69}, 016122 (2004).
\bibitem{H03} H. Hinrichsen, Physica A {\bf 320}, 249 (2003).
\bibitem{BC03} G.T. Barkema and E. Carlon, Phys. Rev. E {\bf 68}, 036113 (2003).
\bibitem{H05} H. Hinrichsen, e-print cond-mat/0501075.
\bibitem{JvWOT04} H.-K. Janssen, F. van Wijland, O. Deloubriere, and
U. C. T\"auber, Phys. Rev. E {\bf 70}, 056114 (2004).
\bibitem{J96} I. Jensen, J. Phys. A {\bf 29}, 7013 (1996).
\bibitem{DL9} See, e.g., I.D. Lawrie and S. Sarbach, in {\it Phase
Transitions and Critical Phenomena}, edited by C. Domb and J.L. Lebowitz
(Academic Press, London, 1984), Vol. 9.
\bibitem{Tri} T. Ohtsuki and T. Keyes, Phys. Rev. A {\bf 35}, 2697 (1987);
{\bf 36}, 4434 (1987); S. L\"ubeck (unpublished); P. Grassberger,
e-print cond-mat/0510428.
\bibitem{comment} Our definition of the crossover exponent is the inverse
of that in Ref. \cite{Tri} and there might be a logarithmic
correction though not visible.
\end{thebibliography}
\end{document}